\documentclass[%
 reprint,
 amsmath,amssymb,
 aps,
]{revtex4-1}

\usepackage{graphicx}
\usepackage{dcolumn}
\usepackage{bm}
\usepackage{braket}

\begin{document}


\title{Comparison of Donor-Acceptor $\pi$-Conjugated Dyes in Model Solar Cells: A Study of Interfacial Ultrafast Electron Migration}

\author{Gunter Hermann}
 \altaffiliation{Institut f\"ur Chemie und Biochemie, Freie Universit\"at Berlin, Takustra{\ss}e 3, 14195 Berlin, Germany}
\author{Felix Witte}%
 \altaffiliation{Institut f\"ur Chemie und Biochemie, Freie Universit\"at Berlin, Takustra{\ss}e 3, 14195 Berlin, Germany}
\author{Jean Christophe Tremblay}
 \altaffiliation{Institut f\"ur Chemie und Biochemie, Freie Universit\"at Berlin, Takustra{\ss}e 3, 14195 Berlin, Germany}

\date{\today}

\begin{abstract}
Interfacial ultrafast electron migration processes are simulated in finite cluster models of
dye-sensitized solar cells within a single active electron approach.
Initially, three different donor-acceptor $\pi$-conjugated dyes supported on colloidal 
titania clusters are compared from the perspective of their optical and electronic properties.
The potential performance of the model solar cell devices for charge migration 
processes is predicted from a static perspective.
For this purpose, parameter-free expressions for state-resolved injection times 
and currents are established and evaluated for the three systems.
A broadband laser excitation promoting the excited states in the visible region
initiates the electron migration process.
The evolution of the electronic wave packet is analyzed with a density-based toolset 
including the electronic yields partitioned for characteristic fragments of the model complexes 
and the time-dependent one-electron density for distinctive time steps in the dynamics.
On the one hand, these reveal a microscopic picture for the mechanistic pathway of 
the charge migration and on the other hand, they validate the results from the time-independent
analysis concerning the photovoltaic efficiency.
\end{abstract}

\maketitle


\section{Introduction\label{intro}}

Dye sensitized solar cells (DSSC), also known as Gr{\"a}tzel cells,
are promising candidates to satisfy in the long-term the steadily growing global 
energy demand for an environmentally sustainable energy source.\cite{Gratzel1991low,hagfeldt1995light,hagfeldt2000molecular,gratzel2001photoelectrochemical,asbury2001ultrafast,gratzel2003dye,hagfeldt2010dye,dyes} 
This prospect is substantiated by their ability to capture and directly convert 
solar light into electrical current with high efficiency.\cite{efficient,efficient2,efficient3,efficient4}
That in turn qualifies them as potential alternatives to conventional silicon-based 
photovoltaic devices.
In general, these Gr{\"a}tzel type solar cells assemble transition-metal or 
organic dye molecules with a wide band gap semiconductor, such as titanium 
dioxide (TiO$_2$).
This functionalization of semiconductor surfaces influences their electronic and 
optical properties and enables DSSCs the harvesting of sun light.
The photonic energy conversion in DSSCs starts with the adsorption of incident light.
This initiates the excitation of the dye from its ground state to an 
excited state. 
The latter is energetically located in the semiconductor conduction band.
The photo-excited electron is then transferred to the semiconductor substrate on 
an ultrafast time scale typically ranging from femto- to picoseconds.\cite{schnadt2002experimental,huber2002real}
Immediately, a charge separation occurs at the interface between the dye and the substrate.
Ensuingly, the electron flows to one of the electrodes, and the oxidized dye is regenerated 
by the redox mediator, such as [Cu(dmp)$_2$]$^{+1/+2}$\cite{mediator} or TEMPO\cite{mediator2}, 
in the surrounding solution.
In turn, this redox pair is restored at the counter electrode.
The photoinduced electron-hole pair separation is the key step for the direct conversion 
of sunlight into electrical energy in DSSCs and thus influences  its performance.
The efficiency of this process strongly depends on the structural, optical, and 
electronic properties of the dye/semiconductor system, especially the proper alignment 
of their energy levels.\cite{performance,performance2,performance3}
Consequently, the choice of the dye is crucial for a beneficial surface sensitization 
of the semiconductor.

Besides transition-metal dyes yielding the highest conversion efficiencies,\cite{chiba2006dye,mathew2014dye}
the burgeoning group of metal-free organic chromophores have attracted particular interest.
They stand out by their tunable optical properties, reduced environmental impact,
low-cost, and rather simple synthetic routes.\cite{hagberg2008molecular,ooyama2009molecular,mishra2009metal,pastore2010computational}
In particular, donor-acceptor $\pi$-conjugated photosensitizers based on
pyridinium are promising representatives.\cite{fortage2010designing,fortage2010expanded,cheng2012dye}
They are based on an expanded $\pi$-conjugated moiety with a pyridinium core as electron acceptor, 
an amino group as electron donor, and a carboxylate group anchored  
to the semiconductor substrate.
Its high photovoltaic performance relies on the broad and intense sunlight absorption 
of the $\pi$-conjugated system and the efficient charge separation in the dye due to the
wide spatial separation of the electron releasing and electron withdrawing groups.\cite{le2011theoretical,ciofini2013JPCL}

A myriad of theoretical\cite{theostudy,theostudy2,theostudy3,theostudy4,yang2015computational,Gomez2017215} 
and experimental\cite{expstudy,expstudy2,expstudy3,expstudy4} studies has been conducted in the recent years 
with particular focus on the characterization of the spectroscopic and electronic properties
of a multitude of different DSSCs and on the examination of the ultrafast charge injection.\cite{duncan2007theoretical,martsinovich2011theoretical}
All of these follow the general aim of gaining information about the potential photovoltaic 
performance of the different chromophores supported on a TiO$_2$ substrate.
In this context, the light absorption, the energetic alignment, and the electron injection 
times of DSSCs are pivotal properties which influence their photon-to-current conversion
efficiency.
From a theoretical point of view, these characteristics are chiefly extracted from static electronic structure 
calculations at the density functional theory (DFT) level, and from absorption spectra 
from linear-response time-dependent DFT (TDDFT).\cite{le2009td,zarate2016nature}
Within a single particle approach, we are aiming to ascertain these properties using finite 
cluster models in a comparative study for three of these so-called push-pull dyes 
attached to a colloidal TiO$_2$ substrate.
This approach has been proven to be reliable to describe such solar cell complexes.\cite{gomez2015imaging,hermann2015laser}
In this framework, we advocate calculating absorption spectra and projected density of states (pDOS) 
and elaborating parameter-free expressions for the state-resolved injection times and currents.
In addition, we put special emphasis on time-resolved simulations of ultrafast interfacial 
charge migration processes in the model devices.
These are described by time-dependent one-electron equations of motion and
are initiated with a broadband laser excitation.
In order to unveil the mechanistic features of the electron dynamics, we rely on a 
density-based diagnostic toolset which has demonstrated its suitability in previous 
studies of charge migration processes.\cite{gomez2015imaging,hermann2016ultrafast,ci_orbkit_wf}
This toolset includes the projection of the electronic yield onto characteristic molecular 
fragments of the DSSCs and time-dependent electron densities.
These highlight the correlation between the structural features of the dyes 
and the mechanistic pathway of the charge migration process.

The remainder of this paper is structured as follows: 
The composition of the model systems and the theoretical methodology to characterize and analyze the
properties of the solar cell devices and to simulate the ultrafast charge migration process 
are outlined in Sec. \ref{theory}.
The subsequent section presents the static characterization of the model complexes and
the laser-driven electron dynamics revealing the system's photovoltaic performance.
The most significant findings are summarized in the last section.

\section{Model Systems and Theory\label{theory}}

In this work, the light-induced interfacial electron migration processes in dye-semiconductor
solar cell systems are simulated in real time and at an atomistic level of theory.
The essential prerequisites to perform and analyze the dynamical simulation as 
well as the studied model solar cells are described below.

\subsection{Construction of the Model Systems\label{model}} 

\begin{figure*}
  \centering
(a) \includegraphics[width=0.25\textwidth]{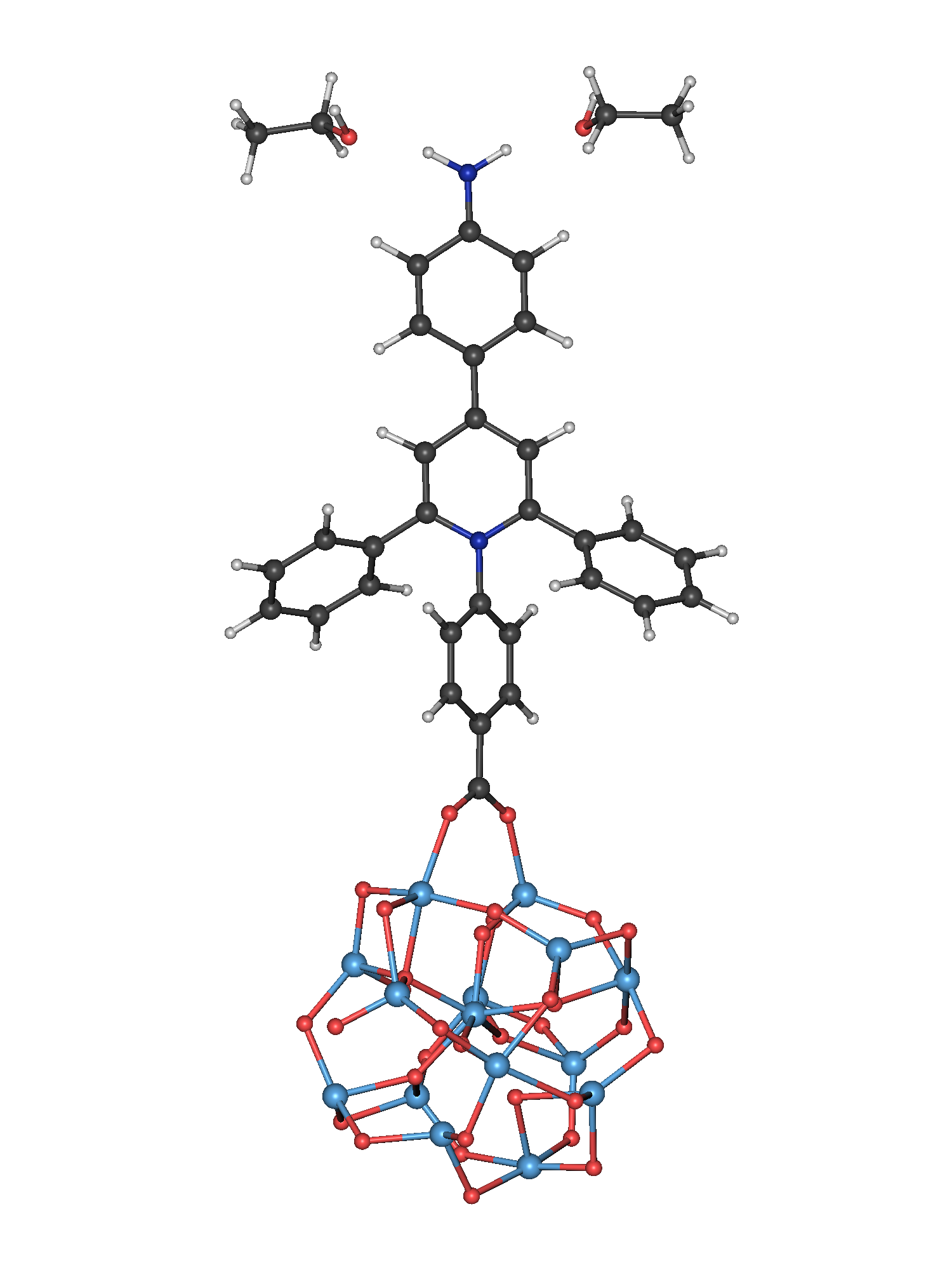}
(b) \includegraphics[width=0.25\textwidth]{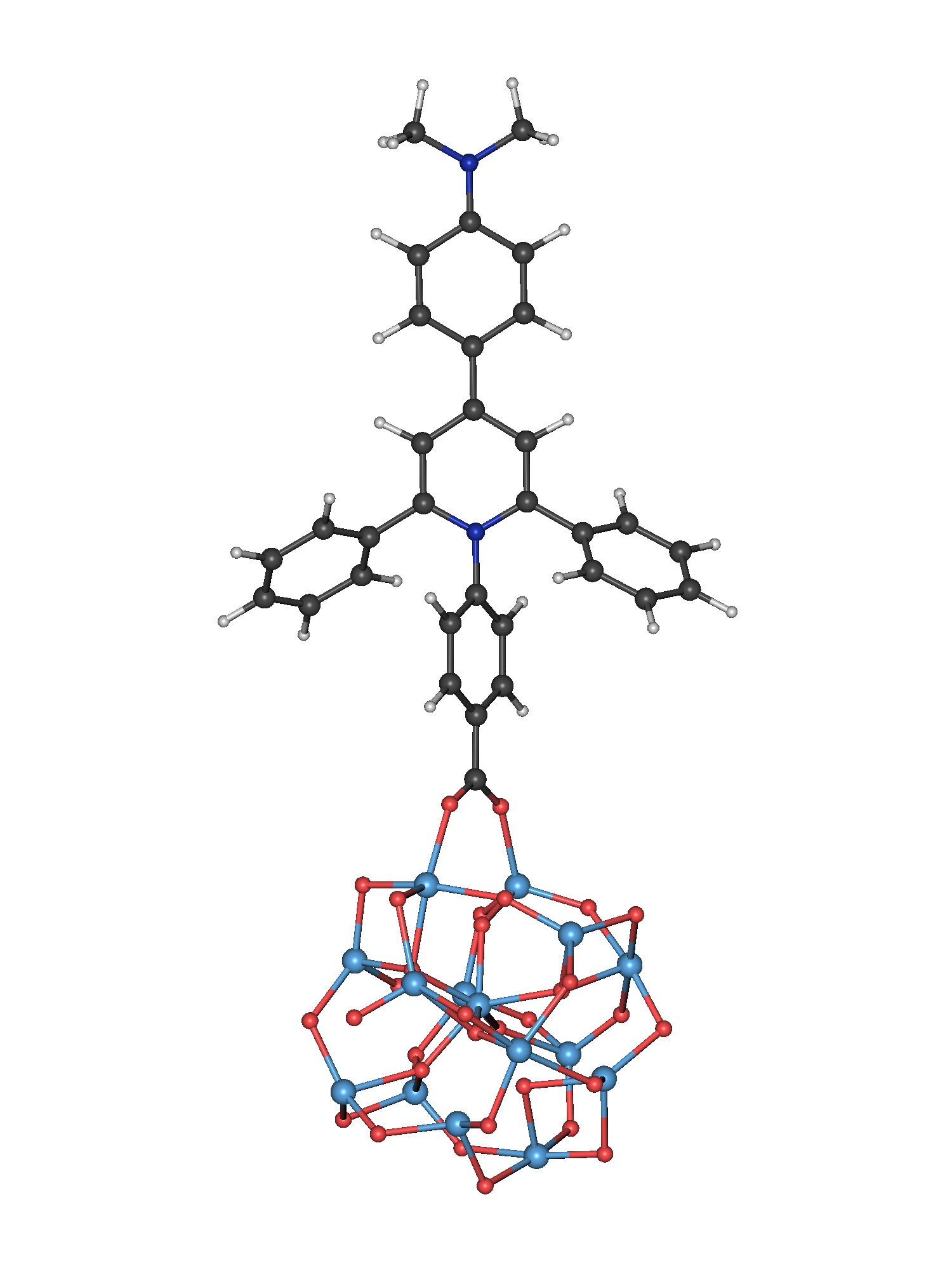}
(c) \includegraphics[width=0.25\textwidth]{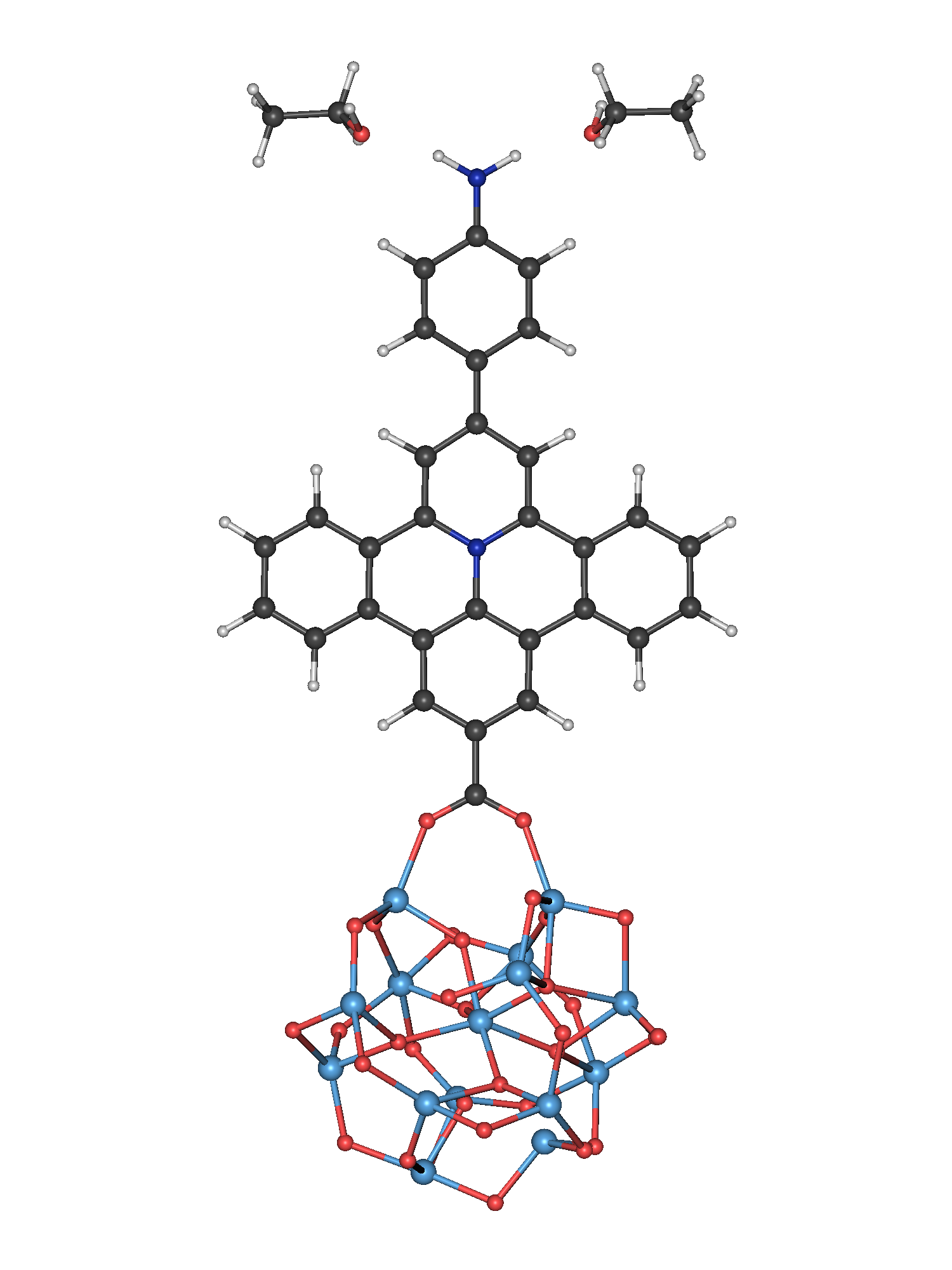}
\caption{Ball-and-stick models of the optimized molecular structures of the 
three different dyes, (a) \textbf{B1}, (b) \textbf{B2}, and (c) \textbf{F1}, attached 
to a colloidal TiO$_2$ nanocluster. 
The carbon, hydrogen, nitrogen, oxygen, and titanium atoms are represented as dark gray, 
light gray, blue, red, and cyan beads, respectively.
}
\label{structure}
\end{figure*}

The model systems for the dye-sensitized solar cells are constructed as finite cluster 
models, where a chromophore is attached to a colloidal TiO$_2$ nanocluster.
Three different organic dyes are investigated concerning their performance and 
efficiency as photosensitizers.
These three belong to the family of donor-acceptor $\pi$-conjugated dyes possessing 
a similar architecture.
The common structural features of these push-pull dyes comprise an electron releasing  
group and an electron withdrawing group.
In the present dyes, the latter is a conjugated $\pi$-system consisting of a pyridinium 
core expanded by three benzene rings.
Two of the dyes (\textbf{B1} and \textbf{B2}) exhibit a branched structure for the electron acceptor moiety,
where the benzene rings are not directly bound to each other.
In the third dye (\textbf{F1}), the rings and the pyridinium core are fused to a flat, rigid, and strongly 
conjugated $\pi$-system.
The respective nomenclature, i.e., \textbf{B} for branched and \textbf{F} for fused, 
stems from Ciofini and coworkers, who have previously investigated these dyes from a 
purely static point of view.\cite{ciofini2013JPCL}
Their findings have revealed advantageous electrochemical and optical properties of
these dyes.
As electron donor group, two different types of amino groups are investigated: 
a primary amino group (-NH$_2$) for \textbf{B1} and \textbf{F1} and a dimethyl amino group (-NMe$_2$) for \textbf{B2}.
In the model solar cells, the dyes are anchored to the colloidal TiO$_2$ nanoparticle
via a carboxylate group with a bidentate bridging attachment.
This binding mode is typical for carboxylate groups resulting in a relatively strong 
chromophore-semiconductor coupling.\cite{carboxylateanchor,carboxylateanchor2}
The titania nanostructure is composed of 15 TiO$_2$ units and corresponds to a 
free-standing nanoparticle in solution.
The starting point for its optimized geometry is a spherically shaped particle suspended 
in a polar solvent, ethanol.
Its cluster size is chosen in accordance with the work of S{\'a}nchez-de-Armas \textit{et al.}\cite{sanchez2010real}, 
who demonstrated that this moderate sized nanocrystallite is sufficient to recover the 
electronic and optical properties of experimentally investigated dye-sensitized solar cells.
The optimized geometry of all three model solar cell devices is depicted in Fig.~\ref{structure}.

\subsection{Dynamical Simulation}

In this work, the light-induced electron migration processes in dye-sensitized solar 
cells are described within a time-dependent single active electron approach.
The real-time propagation of the electron dynamics is performed by solving the 
one-particle non-relativistic time-dependent Schr{\"o}dinger equation\cite{schrodinger1926quantisierung}
\begin{equation}\label{TDSE}
  \imath \hbar \frac{\partial}{\partial t} \Psi_{\textrm{el}}\left(\mathbf{r}, t\right) 
  = \hat{H}\left(\mathbf{r},t\right) \Psi_{\textrm{el}}\left(\mathbf{r},t\right).
\end{equation}
Here, $\Psi_{\textrm{el}}\left(\mathbf{r}, t\right)$ is the time-dependent one-particle
electronic wavefunction, and $\hat{H}\left(\mathbf{r},t\right)$ signifies the total time-dependent
Hamiltonian.
In general, the interfacial electron migration process in a dye-sensitized solar cell 
comprises three essential steps: 
Firstly, the photoexcitation of the chromophore, secondly, the migration of the photoelectron from the dye
into the colloidal semiconductor substrate, and thirdly, the absorption of the electron
at the edges of the titania cluster to simulate the contact with an infinite electrode.
To capture all these steps, the total Hamiltonian is modeled as follows
\begin{equation}\label{TDH}
  \hat{H}\left(\mathbf{r},t\right) = \hat{H}_{0}\left(\mathbf{r}\right) + \hat{H}_{\textrm{field}}\left(t\right) - \hat{W}_{\rm CAP},
\end{equation}
where $\hat{H}_{0}\left(\mathbf{r}\right)$ is the field-free time-independent 
electronic Hamiltonian, $\hat{H}_{\textrm{field}}\left(t\right)$ characterizes the
interaction of the system with a time-dependent field, and $\hat{W}_{\rm CAP}$ is a
complex absorbing potential.

In the present single active electron approach, the evolution of the electronic structure
of the system out of equilibrium is obtained from the time-dependent electronic wave function
which is defined as an expansion in a basis of stationary one-particle functions
\begin{equation}\label{td_wf}
\Psi_{\textrm{el}}\left(\mathbf{r},t\right) = \sum_{\lambda=0} C_{\lambda}\left( t\right) \varphi_{\lambda}\left(\mathbf{r}\right).
\end{equation}
Here, $C_{\lambda}\left( t\right)$ signifies the expansion coefficients of the electronic state $\lambda$,
and $\varphi_{\lambda}\left(\mathbf{r}\right)$ are the respective time-independent one-particle eigenfunctions.
The latter satisfy the stationary one-particle electronic Schr{\"o}dinger equation
\begin{equation}
  \hat{H}_{0}\left(\mathbf{r}\right) \varphi_{\lambda}\left(\mathbf{r}\right) = E_{\lambda} \varphi_{\lambda}\left(\mathbf{r}\right).
\end{equation}
A single occupied orbital describes the initial state, which can be promoted to several
excited orbitals.
This computational setup corresponds to a restricted active space configuration interaction
singles scheme.\cite{foresman1992toward}
To obtain the time-evolution of the coefficients $C_{\lambda}\left( t\right)$,
the equations of motion are integrated numerically using an adaptive step size
Runge-Kutta algorithm, as implemented in our in-house program package $\rho$-TDCI.
A detailed description can be found elsewhere.\cite{tremblay2004using,tremblay2008time,tremblay2011dissipative}

The eigenfunctions $\varphi_{\lambda}\left(\mathbf{r}\right)$ are derived from first-principles electronic 
structure calculations at the DFT level and thus coincide with Kohn-Sham (KS) orbitals $\varphi_{\lambda}\left(\mathbf{r}\right)$.
In the MO-LCAO (Molecular Orbital-Linear Combination of Atomic Orbitals) 
ansatz, these KS orbitals are expanded as a linear 
combination of a finite set of atomic orbitals $\psi_{i}\left(\mathbf{r}-\mathbf{R}_{A}\right)$
\begin{equation}\label{MOLCAO}
\varphi_{\lambda}\left(\mathbf{r}\right)=
\sum_{A}^{N_{\textrm{A}}}\sum_{i}^{N_{\textrm{AO}}}D_{\lambda i}\psi_{i}\left(\mathbf{r}-\mathbf{R}_{A}\right),
\end{equation}
where $D_{\lambda i}$ is the $i$th MO coefficient for the MO $\lambda$, 
$N_{\textrm{A}}$ denotes the number of atoms, and $N_{\textrm{AO}}$ stands for 
the number of atomic orbitals.
In this approach, the energies of the eigenstates (cf. Eq. \ref{td_wf}) correspond
to the KS orbital energies, and the Highest-Occupied MO (HOMO) is typically selected 
as the orbital containing the active electron.
Since the HOMO is usually located on the dye, the photoexcitation in DSSCs is dominantly driven 
by intramolecular transitions between the HOMO and several virtual orbitals.
Thus, an appropriate selection of virtual KS orbitals is required to properly simulate
the excitation dynamics.
In accordance to the usage of KS orbitals, the time-dependent single active electron
ansatz is referred to as TDKS.
The excitation dominated by a HOMO-LUMO character obtained from linear-response TDDFT is used to
scale the fundamental gap of the KS orbitals.

In order to photoexcite the DSSC model system, the semiclassical dipole approximation
is used for $\hat{H}_{\textrm{field}}\left(t\right)$
\begin{equation}
 \hat{H}_{\textrm{field}}\left(t\right) = - \boldsymbol{\hat{\mu}}\cdot\mathbf{F} \left( t\right)
\end{equation}
with $\boldsymbol{\hat{\mu}}$ as molecular dipole operator.
A superposition of linearly polarized laser fields $\mathbf{F} \left( t\right)$  
with a $\sin^2$-shape
\begin{eqnarray}\label{laser}
  \mathbf{F} \left( t\right) & = & \sum_{k} \mathbf{f}_{k}\left( t\right) \cos \left(\omega_{k} t\right),\nonumber\\ \\
  \mathbf{f}_{k}\left( t\right) & = & \begin{cases}\mathbf{f}_{k,0} \sin^2 \left( \frac{\pi t}{t_{k,f}} \right) & {\textrm{if}} \quad 0 < t < t_{k,f} \\ 0 \quad &  \textrm{else}\,, \end{cases}\nonumber
\end{eqnarray}
is applied to promote several excited states from the initial state.
In all simulations, the system is initially found in the ground state, which is represented
here by the HOMO.
In Eq. \ref{laser}, the carrier frequency of the laser pulse $k$, its pulse length, and 
its amplitude are designated by $\omega_{k}$, $t_{k,f}$, and $\mathbf{f}_{k,0}$, respectively.
Here, the carrier frequencies correspond to the difference energy between the ground state
and the target states.

Since the electron migration process is simulated in a finite Ti$_2$O nanocrystallite, 
a complex absorbing potential 
$\hat{W}_{\textrm{CAP}}=\sum_{\lambda} \gamma_{\lambda} \Ket{\varphi_{\lambda}} \Bra{\varphi_{\lambda}}$ 
(cf. Eq. \ref{TDH}) is employed to suppresses artificial recurrences in the electron 
dynamics and reflections at the boundary of the model DSSC.
To this end, we partition the time-independent Hamiltonian as
\begin{eqnarray}
\hat{H}_0 & = & \left(\hat{P}+\hat{Q}\right)\hat{H}_0\left(\hat{P}+\hat{Q}\right)\nonumber\\
& = & \hat{P}\hat{H}_0\hat{P} + \left(\hat{P}\hat{H}_0\hat{Q} + \hat{Q}\hat{H}_0\hat{P}\right) +\hat{Q}\hat{H}_0\hat{Q},
\end{eqnarray}
where $\hat{P}$ is the Mulliken projector on the interfacial oxygen atoms, and
$\hat{Q}=1-\hat{P}$ is its complement. 
We define the absorbing potential as the hopping term from the dye-cluster system to
the interface, where the electron is absorbed at the rate $\gamma_{\lambda}$.
The associated intrinsic lifetime $\tau_{\lambda} = \gamma_{\lambda}^{-1}$ 
of state $\lambda$ can be evaluated using first order time-dependent perturbation theory
\begin{eqnarray}\label{ap_hopping}
 \gamma_{\lambda} & = & \frac{\pi}{\hbar} \left| \Braket{\varphi_{\lambda} \left| \hat{W}_{\textrm{CAP}} \right| \varphi_{\lambda}} \right|^2 \delta \left(E_{\lambda}-E_{\lambda}\right) \nonumber \\
  & = & \frac{\pi}{\hbar} \left| \Braket{\varphi_{\lambda} \left| \hat{P}\hat{H_0}\hat{Q} + \hat{Q}\hat{H_0}\hat{P}\right| \varphi_{\lambda}} \right|^2 \delta \left(E_{\lambda}-E_{\lambda}\right) \nonumber\\
  & = & \frac{\pi}{\hbar} \Bigg| \sum_{\kappa\kappa'} \Braket{\varphi_{\lambda} \left|\hat{P} \right| \varphi_{\kappa}} \Braket{\varphi_{\kappa} \left|\hat{H_0} \right| \varphi_{\kappa'}}
  \Braket{\varphi_{\kappa'} \left|\hat{Q} \right| \varphi_{\lambda}} \nonumber \\
  & & + {\rm c.c.} \Bigg|^2 \delta \left(E_{\lambda}-E_{\lambda}\right) \nonumber \\
  & = & \frac{\pi}{\hbar} \Bigg| \sum_{\kappa} \Braket{\varphi_{\lambda} \left|\hat{P} \right| \varphi_{\kappa}} E_{\kappa}
  \Braket{\varphi_{\kappa} \left|\hat{Q} \right| \varphi_{\lambda}} \nonumber \\
  & & + {\rm c.c.} \Bigg|^2 \delta \left(E_{\lambda}-E_{\lambda}\right).
\end{eqnarray}
Evaluation of the matrix element $\Braket{\varphi_{\lambda}|\hat{P}|\varphi_{\kappa}}$
from the interfacial projector formalism\cite{gomez2015imaging,li2010theoretical} is 
performed as
\begin{equation}\label{ap_proj}
\Braket{\varphi_{\lambda}|\hat{P}|\varphi_{\kappa}}
=  \sum_{A\in \textrm{O}}^{N_{\textrm{A}}}\sum_{i}^{N_{\textrm{AO}}}\sum_{j_{A}}^{N_{\textrm{AO}}}C_{\lambda j_{A}}C_{\kappa i}S_{j_{A}i},
\end{equation}
where $S_{j_{A}i}=\braket{\psi_{j_{A}}|\psi_{i}}$ is an element of the atomic 
orbital overlap matrix.
The adsorption projector $\hat{P}$ exclusively acts on the oxygen atoms 
at the bottom of the titania cluster of the model DSSC.

\subsection{Static and Dynamical Analysis}

In the scope of the present work, two main goals are pursued: first, the characterization
of the static electronic and optical properties for our DSSC model systems, and
second, the analysis and visualization of the dynamical mechanism of the photo-induced
electron migration processes.
For the first, the atom-projected density of states (pDOS) and the absorption spectra
of the model DSSCs are calculated from the KS orbitals to verify the proper energetic 
alignment between the dye and the TiO$_2$ cluster and to validate their spectroscopic
properties.
For this purpose, the excitation energies $\Delta E_{0 \lambda}$ and the oscillator strength 
$f_{0\lambda}$ between the HOMO and a set of target virtual orbitals are necessary.
The latter is defined as
\begin{equation}\label{osc_str}
 f_{0\lambda} = \frac{2}{3} \frac{m_{\rm e}}{e^2\hbar^2}\Delta E_{0 \lambda} \sum_{q=x,y,z} \left| \mu_{0\lambda,q}\right|^2
\end{equation}
with $\mu_{0\lambda,q}$ as the $q$th component of the dipole moment.
In our laser-induced electron dynamics, the oscillation strength can be understood 
as the probability of an electronic transition, induced by the interaction with an 
absorbed photon, and the injection time as the time for the injection of the excited 
electron into the semiconductor of the DSSC.
In addition, the current $I_{0\lambda}$ for this electronic state-to-state transition
can be formulated from the oscillator strength $f_{0\lambda}$ and the absorption rate $\gamma_{\lambda}$
\begin{equation}\label{current}
 I_{0\lambda} = e f_{0\lambda} \gamma_{\lambda}.
\end{equation}
Together, the injection time and the current, facilitates to assess the potential performance 
and efficiency of the electron injection process from a time-independent perspective.

The analysis of the real-time propagation of the electron migration processes is 
accomplished by means of a recently presented toolset.\cite{gomez2015imaging,hermann2016ultrafast}
As an essential component, this comprises the time-dependent electron density
\begin{equation}\label{rho}
\rho \left(\mathbf{r},t \right) = \Psi_{\textrm{el}}\left(\mathbf{r}, t\right) \Psi_{\textrm{el}}^{\dagger}\left(\mathbf{r}, t\right).
\end{equation}
For a space-resolved analysis of the time-dependent electron density, the
Voronoi partitioning scheme is employed.\cite{fonseca2004voronoi}
This method assigns certain atoms or fragments to a set of points in space which are 
closest to them.
Integrating the electron density over these Voronoi cells reveals the number of electrons
in the respective spatial region
\begin{equation}\label{enum}
n_{V} \left(t\right) =  \underset{\textrm{cell} \, \textrm{of}\, V}{\underset{\textrm{Voronoi}}{\int}} 
\textrm{d}\mathbf{r} \rho \left(\mathbf{r},t \right).
\end{equation}
Based on the electron density and the Voronoi scheme, one can define a supplementary 
quantity to study the electron dynamics, i.e., the electronic yield
\begin{equation}\label{yield}
Y_{V} \left(t \right) = \int_{0}^{t} \textrm{d}t' \underset{\textrm{cell} \, \textrm{of}\, V}{\underset{\textrm{Voronoi}}{\int}} 
\textrm{d}\mathbf{r} \frac{\partial \rho \left(\mathbf{r},t' \right)}{\partial t'} = n_{V} \left(t\right)-n_{V}\left(0\right).
\end{equation}
This quantity determines the number of electrons that flow through a certain Voronoi cell within a
time interval.\cite{barth2009concerted}

\section{Results and Discussion\label{results}}

\begin{figure*}[t]
\centering
\includegraphics{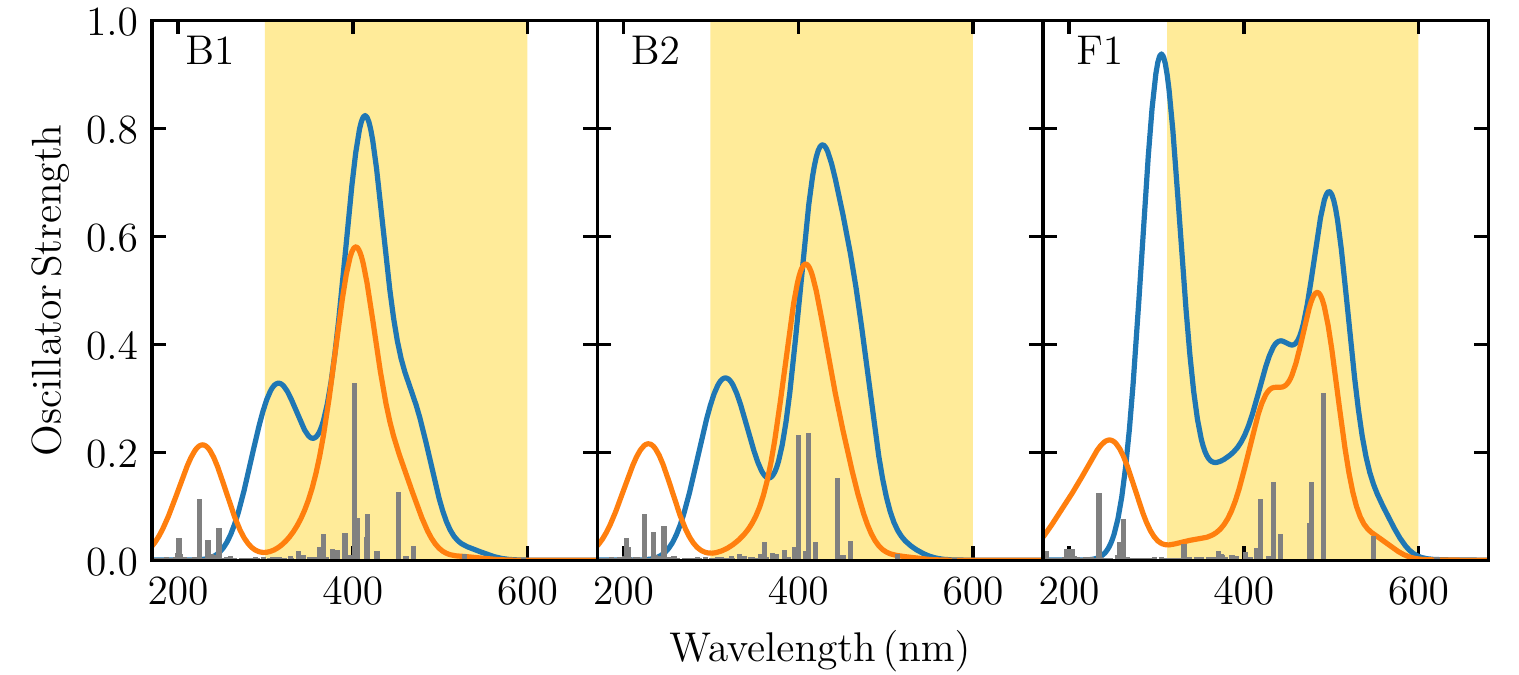}
\caption{Simulated optical spectra for the three extended pyridinium dyes, \textbf{B1}, 
\textbf{B2}, and \textbf{F1}, attached to a colloidal TiO$_2$-cluster. 
Absorption spectra (orange) obtained from the broadened molecular orbital transition
energies ($hc/\Delta E_{0\lambda}$) are shown along with the broadened spectra from the results of linear-response 
TDDFT calculations (blue).
Both are based on a single-point calculation at the PBE0 level using a 
def2-SVP basis set.
The  molecular orbital transition energies corresponds to the difference energies between
the HOMO and the virtual orbitals $\Delta E_{0\lambda}$.
Gaussian functions with a width of $\sigma= 30\,{\rm nm}$ are used to construct the 
spectra.
Vertical gray lines represent the oscillator strengths of the specific transitions of 
the orange spectra. 
The oscillator strengths are calculated according to Eq. \ref{osc_str}.
The yellow region marks the first absorption band in each spectra. 
These excited states are stimulated in the electron dynamics.
}
\label{spectra}
\end{figure*}

\subsection{Computational Details}

Based on the structural construction of the model DSSCs explained in Sec. \ref{model},
structure optimizations for the three dyes, \textbf{B1}, \textbf{B2}, and \textbf{F1},
attached to the colloidal (TiO$_2$)$_{15}$ nanocrystallite are carried out at the DFT level
with the PBE0 functional\cite{adamo1999toward} and the def2-SVP basis set\cite{schafer1992fully,weigend2005balanced}.
In addition, Grimme's dispersion correction D3(BJ) is employed to include the influence  
of  dispersion effects.\cite{grimme2010consistent,grimme2011effect}
During the optimization process, the neutral dye-semiconductor complexes were allowed 
to freely relax.
The continuum solvation model implemented in the COSMO-RS package\cite{klamt1995conductor,eckert2002fast,klamt1998refinement,COSMO} 
is used to mimic the suspension of the colloidal nanoparticles in an electrolyte,
i.e., simulating the effects of ethanol as a solvent on the model DSSCs.
In accordance to the work of Ciofini \textit{et al.}, two explicit ethanol are coadsorbed
to the primary amino group (-NH$_2$) in \textbf{B1} and \textbf{F1} to yield
better optical properties.
These are obtained from TDDFT\cite{gross1990time} calculations 
in linear response to produce electronic  absorption  spectra for the DSSC complexes.
As can be seen from previous work on electron charge migration in DSSCs,
the electronic and spectroscopic properties could be accurately predicted with
the particular selection of TDDFT/PBE0.\cite{tddftpbe0,tddftpbe02}
All quantum chemical calculations are performed with TURBOMOLE\cite{TURBOMOLE}.
For the dynamical simulation of the laser-induced charge migration, an in-house 
implementation of the $\rho$-TDCI method\cite{tremblay2008time,tremblay2011dissipative}
is employed with 35 eigenstates in the dynamical basis.
The ensuing static and dynamical analysis of the charge migration are executed with 
the Python toolbox \textsc{ORBKIT}\cite{orbkit} and the data visualization 
with the programs Amira\cite{Amira} and matplotlib\cite{matplotlib}.
\begin{figure}[t!]
\centering
\includegraphics{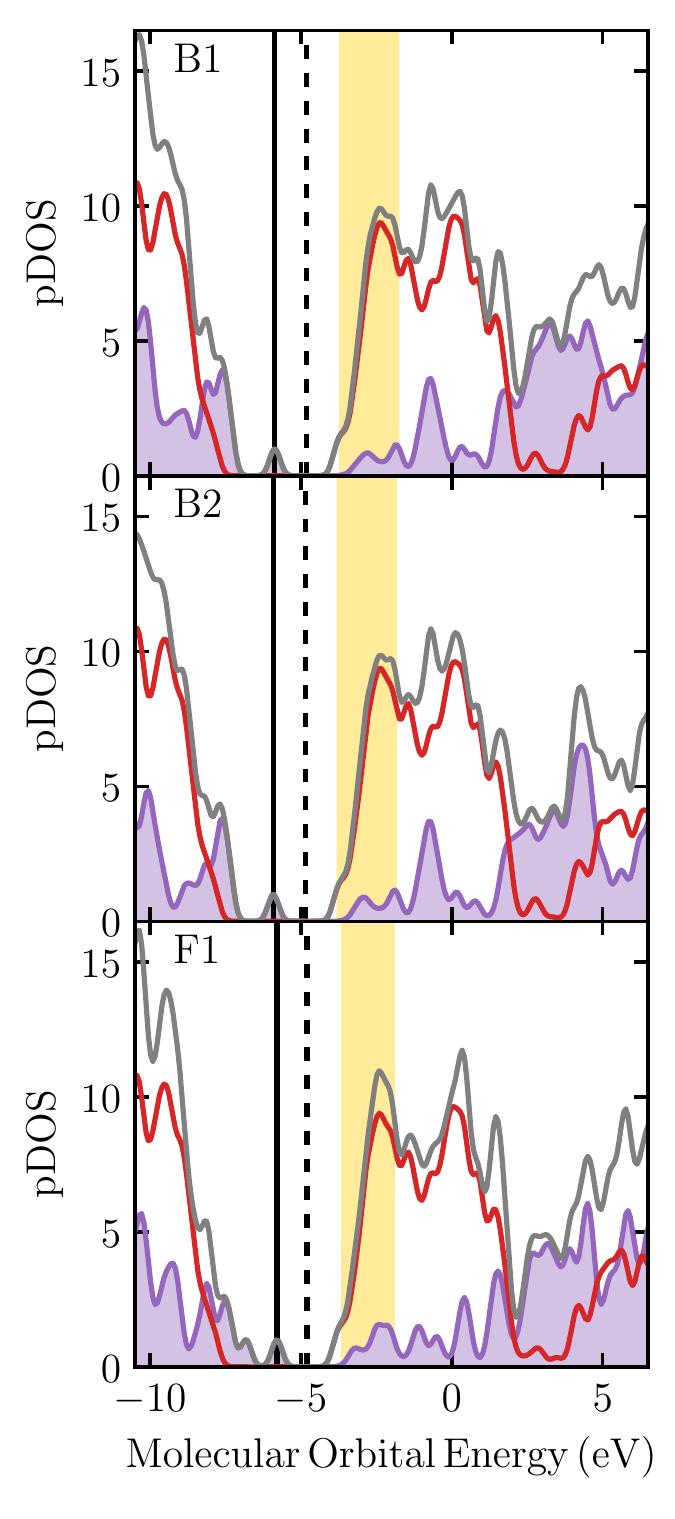}
\caption{Projected density of states (pDOS) for the three photosensitizer, \textbf{B1}, 
\textbf{B2}, and \textbf{F1}, coupled to a titania nanocluster.
The states correspond to the one-electron molecular orbitals.
The DOS is projected on the dye (purple), on the TiO$_2$-cluster (red), and on the entire solar cell
model system (gray).
The solid black line signifies the HOMO and the dashed black line marks the Fermi energy 
(middle of HOMO-LUMO gap).
All excited states which are photoexcited from the HOMO in the electron dynamics are
located in the yellow region.
}
\label{pdos}
\end{figure}
\begin{figure}[t!]
\centering
\includegraphics{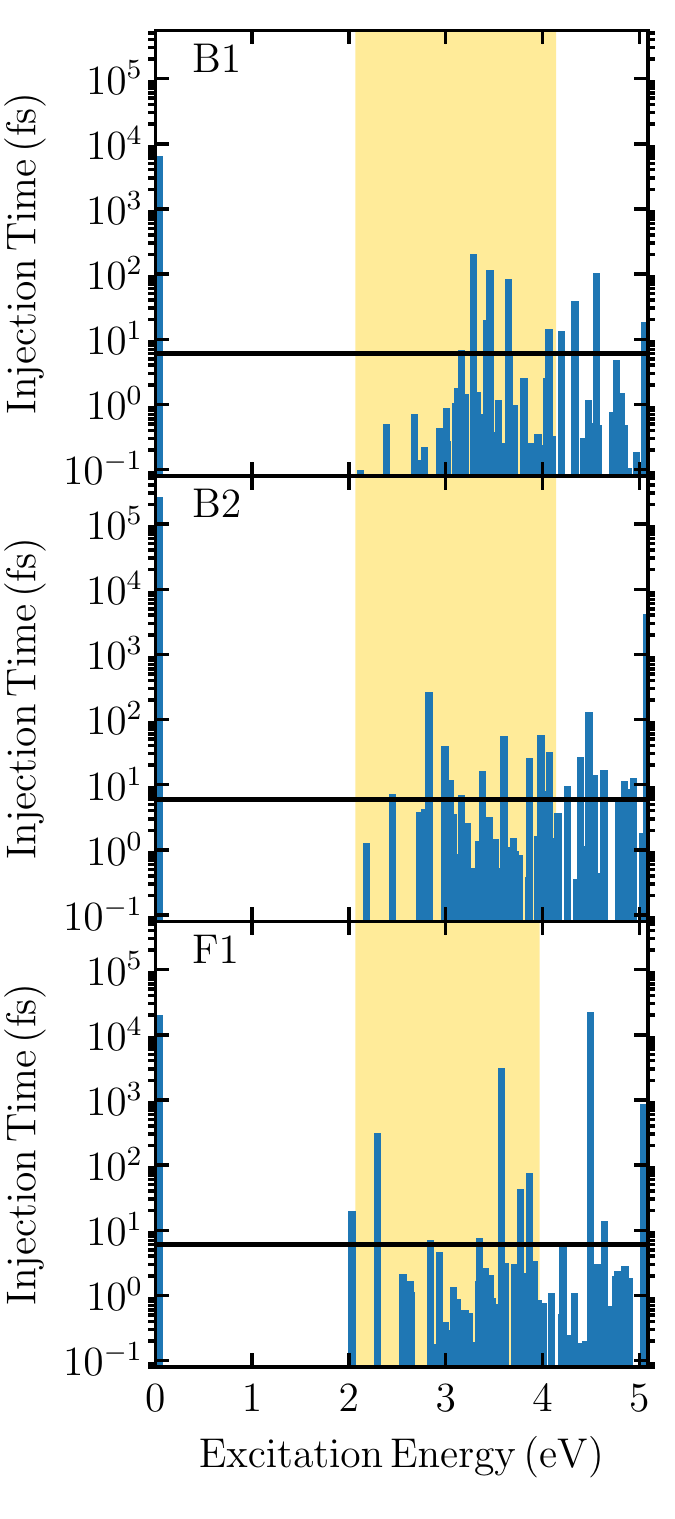}
\caption{Injection time $\tau_{\lambda}$ of the solar cell model systems with \textbf{B1}, \textbf{B2}, and \textbf{F1}, 
as a function of the excitation energy $\Delta E_{0\lambda}$.
The yellow area signifies the region of light excitation in the dynamical propagation.
The black horizontal line at 6 fs signifies a typical experimental injection time for 
similar solar cell devices.
}
\label{injection_time}
\end{figure}

\subsection{Static Properties of the Model Systems}

The optimized structures of the three dye-TiO$_2$ model systems containing the three
different chromophores, \textbf{B1}, \textbf{B2}, and \textbf{F1}, are depicted in 
Fig.~\ref{structure}.
It can be seen in all three complexes that the benzene ring nearest to the titania 
cluster lies in plane with the two anchoring titanium atoms.
In addition, the remaining part of the branched dyes, \textbf{B1} and \textbf{B2}, is twisted 
with respect to this aromatic ring.
The respective dihedral angle between this benzene ring and the pyridinium core is 
$65.3\,{}^{\circ}$ for \textbf{B1} and $66.1\,{}^{\circ}$ for \textbf{B2}.
The reason for this torsion is the reduction of steric hindrance between the benzene ring
nearest to the TiO$_2$-cluster and the two  other freely rotatable benzene rings bound to 
the pyridinium core.
This induces a weaker conjugation of these chromophores in comparison to the fused 
dye, \textbf{F1}.
The pericondensation of the latter dye extends its $\pi$-conjugated system. 
These structural features influence the hybridization between the dye and the 
TiO$_2$ nanocluster, and thus the electronic coupling and energetic alignment between
both.
These are key factors which can crucially affect the charge migration process.
\begin{figure*}
\centering
\includegraphics{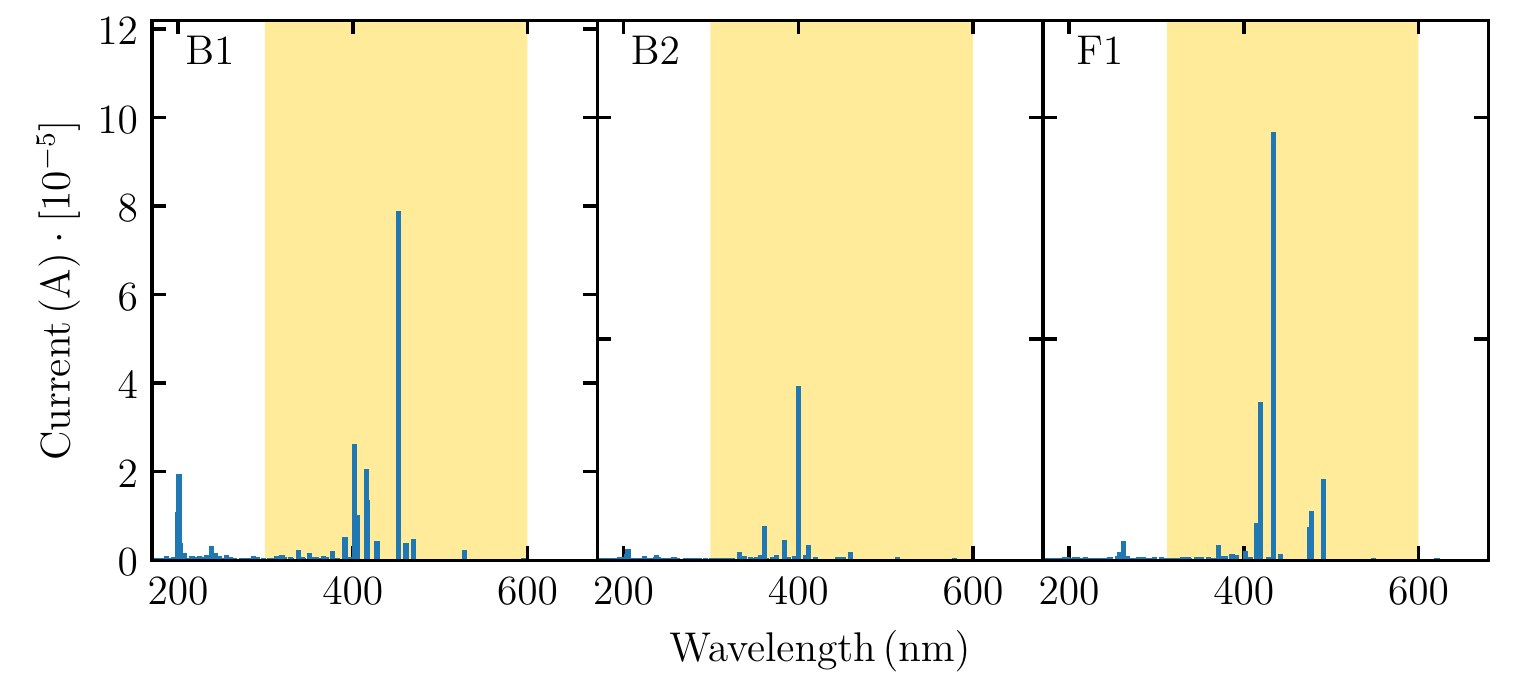}
\caption{Current $I_{0\lambda}$  (cf. Eq. \ref{current}) as a function of the wavelength ($hc/\Delta E_{0\lambda}$) 
for the three studied solar cells.
The yellow area marks the first absorption band of each model system (cf. Fig. \ref{spectra}) 
which are excited in the dynamical simulation.
}
\label{fig_current}
\end{figure*}
Ensuingly, the optical spectra for the three DSSCs model devices are obtained from
the linear-response TDDFT calculations using their PBE0 equilibrium geometries.
For comparison purposes, the absorption spectra are computed on the basis of the MO energy differences
between the HOMO and a set of virtual KS orbitals and the respective oscillator strengths, 
as defined in Eq.~\ref{osc_str}.
Here, the HOMO-LUMO gap is adjusted to the optical gap obtained from the TDDFT results.
A comparison between both methods, TDDFT (blue) and KS (orange), is illustrated in Fig.~\ref{spectra}.
The TDDFT spectra show two broad bands for all three complexes, one in the visual region between 
300 nm and 600 nm, and one in the UV-region.
This shape can be likewise captured with the KS approach. 
From the experimental spectra of the single chromophores (cf. Ref.~\cite{le2011theoretical}), 
a similar shape of the absorption spectra can be recognized, where likewise a broad
band exists in the visual region.
As expected, this band is red-shifted due to the anchoring of the dye to the titania cluster.
In the Supplementary Information, the comparison between both methods with reference to 
the experimental data can be found for the single dyes.
At first glance, the shape of the absorption spectra for the three complexes (cf. Fig.~\ref{spectra}) is 
very similar, while the maxima of the first band is slightly red-shifted from \textbf{B1} 
to \textbf{F1}.
This is a consequence of the different conjugation of the chromophores.
As can be observed, the agreement between both methodologies for the first band is 
satisfactory, revealing a minor blue-shift.
This is due to the fact that the dominant contributions for all excited states of the first
band in the TDDFT reference are excitations from the HOMO to low-lying virtual orbitals.
A density plot of the HOMO can be seen in Fig.~\ref{tdrho} ($t=0\,{\rm fs}$) which 
shows that it is mainly localized on the donor group of each dye.
Consequently, a charge migration from this donor group to the acceptor group and finally 
into the TiO$_2$ cluster is expected during photoexcitation.
The second band is rather poorly reproduced by the KS methodology, 
since transitions from other occupied orbitals except the HOMO are dominant in this
region.
Since excitation of the first band is advocated as photosensitizer in the visible frequency
range, this would not affect the subsequent dynamical simulation.
All states contributing to this band are marked as a yellow shaded area in all figures 
of the static analysis.
To reveal details about the energetic alignment between the dye and the TiO$_2$
nanocluster, the pDOS are presented in Fig.~\ref{pdos} for the dye (purple) and the 
titania cluster (red), together with the pDOS of the complete model device (gray).
In an indirect injection cell, the electron photoexcitation starts from the ground state
localized on the chromophore, which lies energetically in the band gap of the substrate.
The electron is then promoted to an excited state which is hybridized with the conduction band 
of the semiconductor.
As can be seen from Fig.~\ref{pdos}, this energetic alignment is reproduced by the 
single particle eigenstates of all three model DSSCs.
That is, the HOMO marked with a black solid line is fully localized on the 
chromophore and lies between states with dominant contributions from TiO$_2$.
The excited states with optical activity in the visible range are mainly located 
on the substrate, as indicated with the yellow filling.
It is obvious that these states have a slightly higher contribution from the dye 
in the \textbf{F1}-TiO$_2$-complex in comparison to the other two systems, which
implies a higher degree of hybridization between the dye-localized HOMO and the excited 
states in the visual region.
As a partial conclusion, one can state that our model devices correctly mimic the 
essential electronic and spectroscopic properties of a realistic DSSC, which makes 
them suitable for the time-dependent simulation of the electron transfer processes.

In the last step of the time-independent analysis, the possibly achievable photovoltaic 
performance of the three systems is deduced from a static view.
For this purpose, the state-resolved injection times $\tau_{\lambda}$ and the currents $I_{0\lambda}$
for the transition between the HOMO and each one-electron excited state are determined.
The former are depicted in Fig.~\ref{injection_time}, with a reference line corresponding 
to a typical experimental injection time of $6\,{\rm fs}$ (black horizontal line).\cite{huber2002real}
It can be observed that only seven, twelve, and six of the excited states in 
\textbf{B1}, \textbf{B2}, and \textbf{F1} possess an injection time longer than the 
experimental guideline.
The injection time of all other states in the respective region are in the regime of
a few femtoseconds.
Under the assumption that all designated target states can be equally excited from 
the ground state, the advocated charge migration process would proceed fastest for 
the \textbf{F1} dye.

To account for the excitation probability between the HOMO and the virtual orbitals, 
the associated currents $I_{0\lambda}$ are calculated from Eq.~\ref{current}.
Results for the three systems are reported in Fig.~\ref{fig_current}.
For the targeted charge transfer states (yellow region), the total current obtained
by summing all state currents amounts to $1.71\cdot10^{-4}\,{\rm A}$, $5.95\cdot10^{-5}\,{\rm A}$, 
and $1.83\cdot10^{-4}\,{\rm A}$ for \textbf{B1}, \textbf{B2}, and \textbf{F1}, respectively.
It is apparent that the number of charge transfer states dominantly contributing to this 
overall current increases from \textbf{B2} to \textbf{B1} and \textbf{F1}. 
In the charge migration process, these states will be prevalently used by the photoexcited 
electron as channels for the injection from the dye into the TiO$_2$ substrate.
It is interesting to note that one channel at $\sim$450 nm is much more efficient in 
chromophore \textbf{F1}.
This implies that \textbf{F1} is expected to lead to a faster, stronger electron
injection upon photoexcitation.
These tendencies match with the findings of Ciofini and co-workers.

\subsection{Charge Migration Dynamics}

\begin{figure}[t!]
\centering
\includegraphics{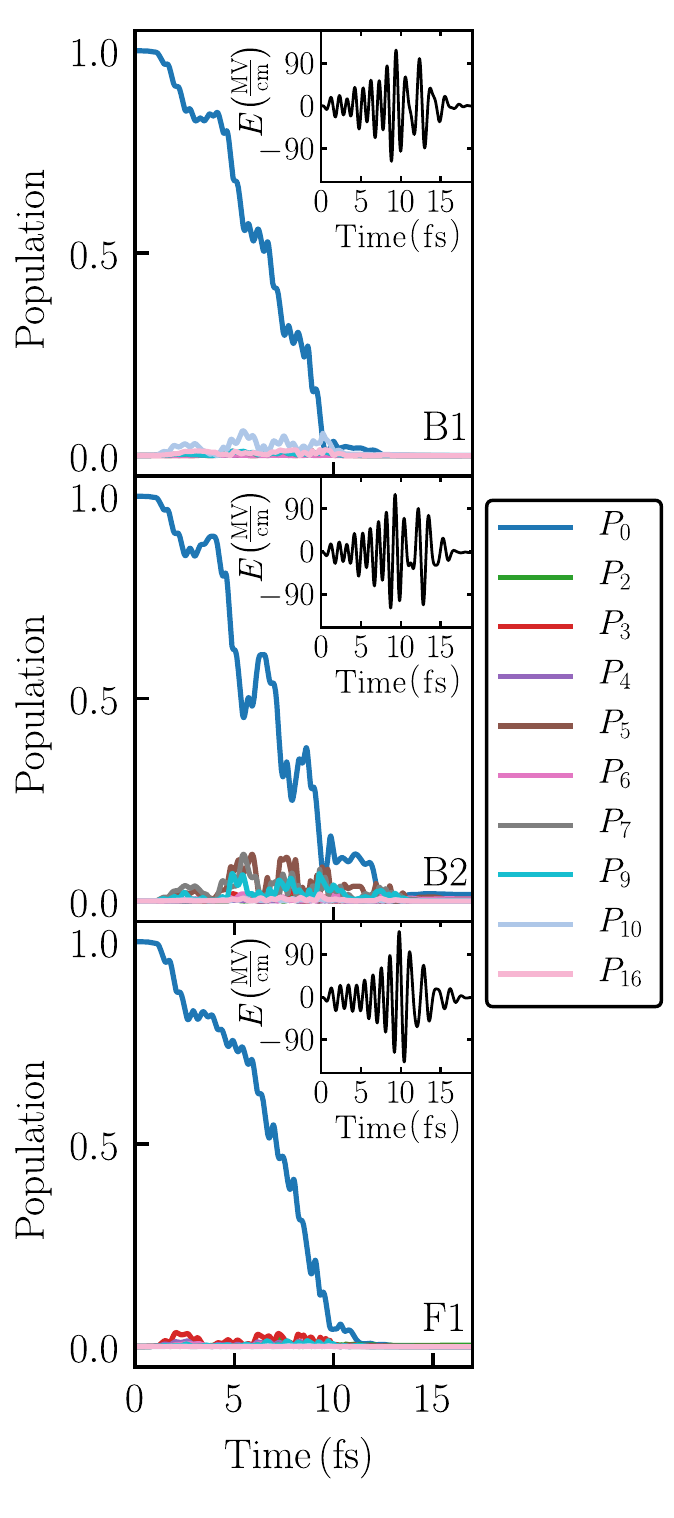}
\caption{Population propagation of the light-induced electron dynamics for the 
three different dyes, \textbf{B1}, \textbf{B2}, and \textbf{F1}, supported on a
TiO$_2$-cluster.
A broadband excitation promotes the electron dynamics.
All target states are located in the first absorption band of each model solar cell
(cf. Fig. \ref{spectra}).
The laser field is constructed from a superposition of state-to-state $\pi$-pulses 
with a duration of 19 fs.
35 states are incorporated in the electron dynamics.
}
\label{pop_evolution}
\end{figure}

The following section examines the differences in the electron transfer dynamics for 
the model solar cells with the three push-pull dyes, \textbf{B1}, \textbf{B2}, and
\textbf{F1}.
In this type of DSSCs, the charge migration is initiated by the photoexcitation of 
the dye, where the electron is transferred from the donor group of the dye 
(amino group) to its acceptor group (extended pyridinium core) and subsequently into the 
semiconductor substrate.
The TiO$_2$ nanocrystallite is in contact with an electrode, which finally absorbs the electron.
To simulate this process, all excited states (virtual orbitals) energetically lying 
in the visible region and building the first absorption band (indicated with a yellow filling in Fig.~\ref{spectra}) 
are excited from the ground state (HOMO) with a superposition of $\sin^2$-shaped pulses.
Each single pulse is tuned to the carrier frequency for the respective state-to-state 
transition.
As for the absorption spectra, the HOMO-LUMO gap is adjusted to the TDDFT optical energy gap.
The pulse amplitude and duration ($19\,{\rm fs}$) are tailored to a typical laser pulse used in experimental studies
of transient absorption measurements in similar DSSCs.\cite{huber2002real,dworak2009ultrafast}
The charge migration process is terminated by the absorption of the electron at the bottom
of the titania nanocluster, as discussed in the previous section.
This is accomplished by the absorption potential defined in Eq.~\ref{ap_hopping} and \ref{ap_proj} 
which concurrently inhibits unphysical reflections at the TiO$_2$ cluster edges.
The propagation of the state populations in real time and the respective laser fields  
for the three systems in comparison are depicted in Fig.~\ref{pop_evolution}.
At first glance, the results for the three solar cells reveal a similar course for the 
photoinduced charge migration.
The ground state is depopulated in the first $\sim$11 fs of the propagation 
during the laser pulse.
The ground state population is transferred to a subset of excited states and 
subsequently absorbed by the potential at the bottom of the titania cluster.
Despite targeting different sets of excited states for the three dyes, the laser fields 
for the broadband excitation have a similar frequency. 
However, the shape differs slightly between the two branched dyes, \textbf{B1} and \textbf{B2},
and the pericondensed dye, \textbf{F1}.
The time-evolution of the state population for the three systems exhibits
distinct characteristics.
On the one hand, the population oscillates differently between the ground state 
and a subset of excited states for the three systems.
On the other hand, the number of significantly populated excited states increases
from \textbf{F1} to \textbf{B2}.
Both have an effect on the overall electron injection time.
The differences can be explained as follows:
The \textbf{B2} dye possesses longer-lived excited states.
Accordingly, the coherent dynamics between these states and the ground state is 
maintained for a longer time period leading to a higher population oscillation.
In contrast, the excited states in \textbf{F1} are very short-lived, and the electron
is rapidly injected into the TiO$_2$-nanocrystallite.
Hence, the excited states are only marginally populated, which results in a lower 
population oscillation with the ground state.
Consequently, the injection process proceeds fastest for \textbf{F1}.

\begin{figure}[t!]
\centering
\includegraphics[width=0.4\textwidth]{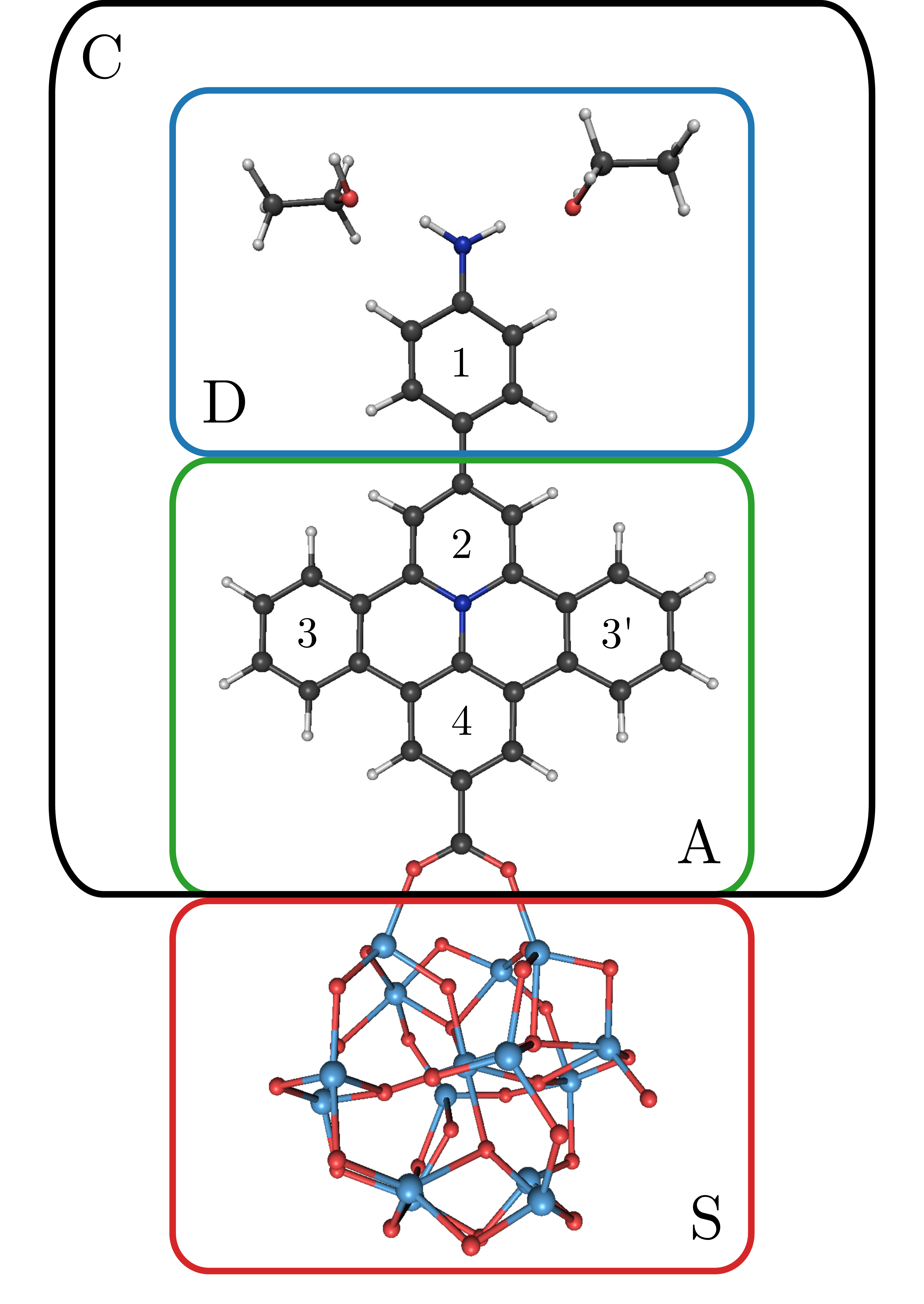}
\caption{Allocation of the Voronoi polyhedrons for the analysis of the electronic yield $Y_{V}\left(t\right)$. 
The boxes determine these particular Voronoi fragments in the respective model 
complexes.
The chromophore (C), its electron donor group (D), its electron acceptor group (A), 
and the TiO$_2$-substrate (S) are colored black, blue, green, and red,
respectively.
Dark gray, light gray, blue, red, and cyan beads represent the carbon, hydrogen, nitrogen, 
oxygen, and titanium atoms, respectively.
}
\label{voro_part}
\end{figure}

\begin{figure*}[t!]
\centering
\includegraphics{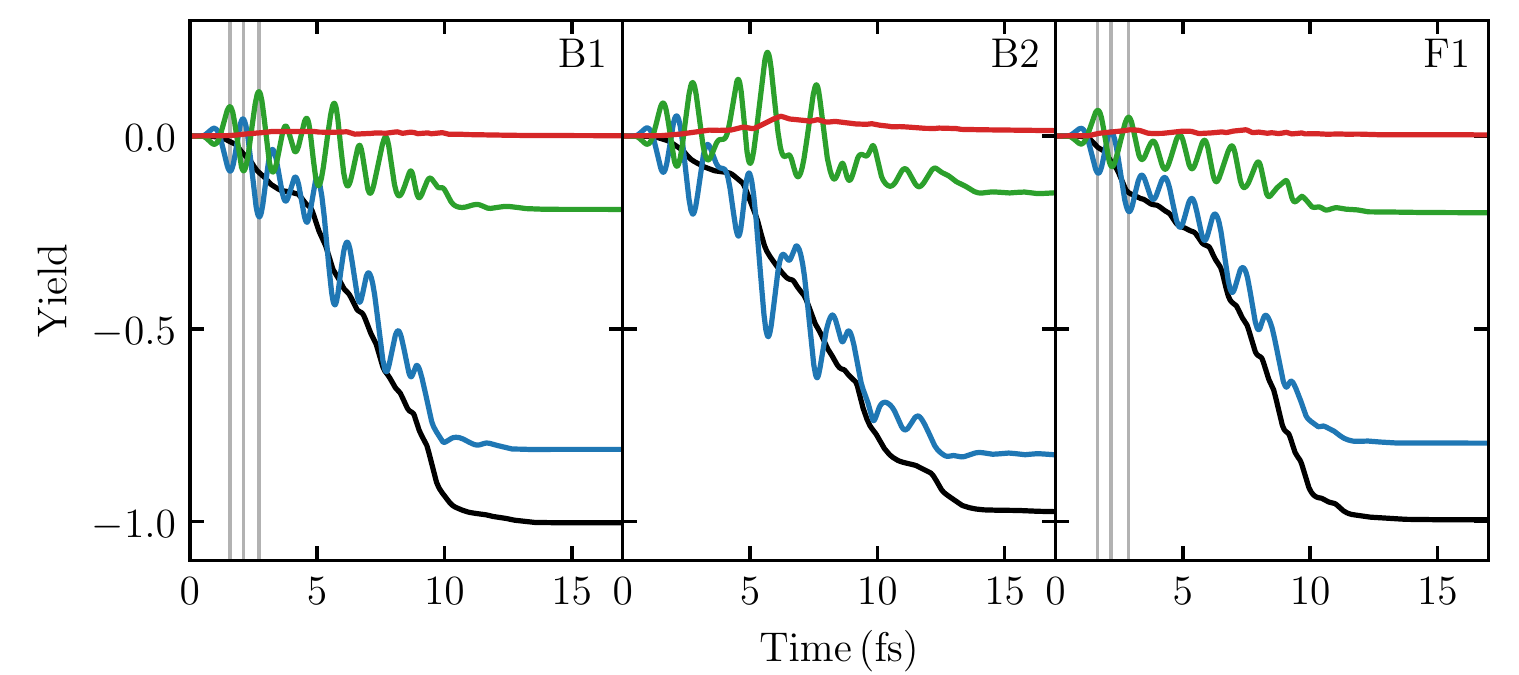}
\caption{Time evolution of the electronic yields $Y_{V} \left(t \right)$ (cf. Eq. \ref{yield}) during 
the charge migration dynamics from the push-pull dyes towards the TiO$_2$ semiconductor.
The yields are calculated for different Voronoi regions comprising the entire dye (black),
its electron releasing (blue) and withdrawing group (green), and the TiO$_2$ nanocluster (red).
}
\label{plt_yield}
\end{figure*}

In order to refine the dynamical analysis, the electronic yields $Y_{V} \left(t \right)$ are calculated from Eq.~\ref{yield}
to reveal the mechanistic details of the electron dynamics.
In relation to the common structural features of the model devices, the yield $Y_{V} \left(t \right)$ is divided 
into certain Voronoi polyhedrons comprising the dye (black, C), its electron donor moiety (blue, D), 
its electron acceptor group (green, A), and the semiconductor substrate (red, S).
This partitioning scheme is represented in Fig.~\ref{voro_part}.
The respective one-electron yields during the laser pulse are displayed in Fig.~\ref{plt_yield}.


Once again, the three solar cell complexes show several similarities: 
First, the electron density synchronously fluctuates between the electron donating and withdrawing 
moieties and is continually injected into the titania nanocluster during this oscillation.
Second, the yield curves for the complete dye and the substrate are relatively smooth in contrast
to the oscillatory pattern of the electron donor and acceptor groups.
Third, the electron density that is injected into the TiO$_2$-cluster is immediately 
absorbed, as can be seen from the almost constant red line.
Here again, dye \textbf{B2} exhibits the most structure because the states are longer-lived
and retain a coherent nature over a longer period of time.
In accordance with the population evolutions, the oscillation of the electronic yields
of the donor and acceptor groups has the largest magnitudes for the branched dyes.
This indicates that the photoexcited electron in the \textbf{F1} dye is transferred 
from the amino group into the semiconductor in a more direct manner.
This can be ascribed to the structural and electronic differences of the dyes.
Considering our static and dynamic analysis, the strength of the charge transfer 
process is very sensitive towards the hybridization between the dye and the TiO$_2$ 
nanocrystallite, which is strongest for the \textbf{F1} dye (cf. Fig.~\ref{pdos}).
Besides, the slightly stronger electron donating property of the dimethyl amino group (-NMe$_2$)
in comparison to the primary amino group (-NH$_2$) appears to decrease the charge 
transfer efficiency.
This observation is in line with the studies of Ooyama and co-workers, who have investigated the 
influence of different substituents in a photosensitizer on the photovoltaic efficiency 
of DSSCs.\cite{ooyama2015effect}

\begin{figure*}[t!]
\centering
\includegraphics[width=1.0\textwidth]{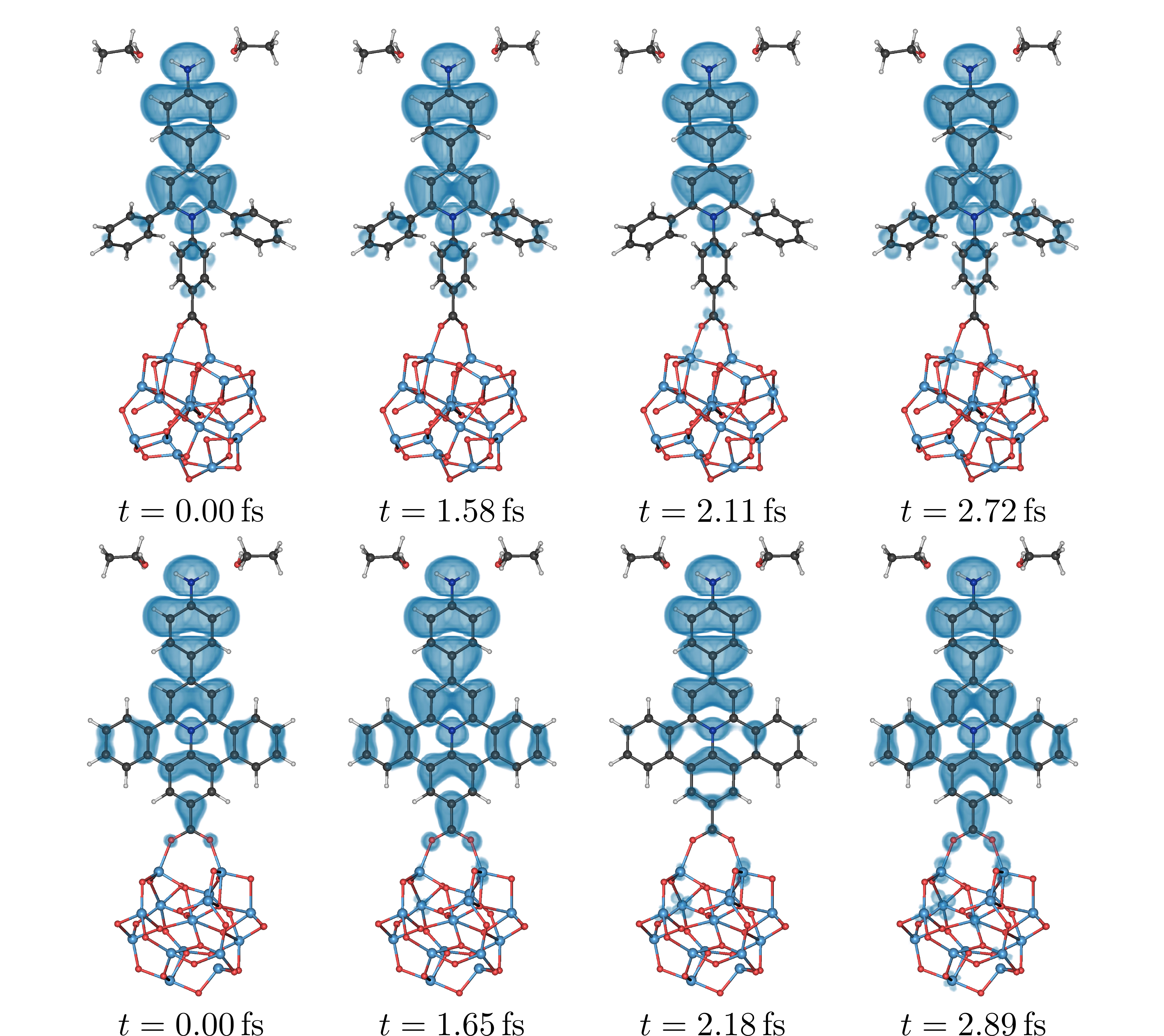}
\caption{
Evolution of the electron density as blue contour plots for the charge migration from the dye to the titania cluster.
A comparison between the structurally branched \textbf{B1} and fused chromophore \textbf{F1} is 
shown for four different time steps. 
These steps are marked in Fig. \ref{plt_yield} with gray vertical lines.
The contour value is set to $10^{-4}$ a$_0^{-3}$.
The carbon, hydrogen, nitrogen, oxygen, and titanium atoms are colored dark gray, 
light gray, blue, red, and cyan, respectively.
}
\label{tdrho}
\end{figure*}

The analysis of the structural properties influencing the charge migration is complemented 
by comparing time-dependent electron densities for chromophore \textbf{B1} 
with a branched framework with the pericondensed dye \textbf{F1}.
Isosurface plots at characteristic time steps are depicted in Fig.~\ref{tdrho}.
The time points are chosen such that the electron density has a maximal time-dependent 
yield at the electron donor group or at the electron acceptor group in order to 
display the fluctuation between both moieties.
Gray vertical lines mark these points in the respective subplots in Fig.~\ref{plt_yield}.
Additionally, the initial time step ($t=0$) is considered.
For sake of clarity, the aromatic rings of the dyes are assigned in Fig.~\ref{voro_part}.
The mechanism of the overall electron injection process can be summarized into two steps:
First, electron density enrichment of the acceptor group (see second and fourth time step in Fig.~\ref{tdrho})
and second, the back transfer of electron density from the acceptor group to the donor 
group and the simultaneous electron injection into the TiO$_2$-cluster (see third time step in Fig.~\ref{tdrho}).
This oscillatory pattern is repeated until all electron density is absorbed at
the bottom of the semiconductor.
In greater detail, it can be observed that the electron moves directly from the amino
group through the benzene rings 2 and 4 to the substrate.
This electron loss appears to be monotonic.
The electron density which is transferred to the rings 3 and 3',
is partially transferred back to the electron releasing group (ring 1).
This effect seems to be increased for the branched dye \textbf{B1}, since there is
no direct injection channel.
In addition, the torsion between ring 4 and the pyridinium core decreases the hybridization
between both fragments, which appears to reduce the injection rate.
Concluding these observations, a large conjugation of the photosensitizer enables 
not only  a broad an intense sunlight  absorption, but also appears to facilitate the electron 
injection.
For the complete picture of the dynamics, a short film of the charge migration process 
is made available online in the Supplementary Information.

\section{Conclusions}

This contribution is a comparative study analyzing the performance and efficiency of 
three donor-acceptor $\pi$-conjugated dyes as photosensitizers in model solar cells.
These chromophores differ by several structural features including two dyes, \textbf{B1} and \textbf{B2}, 
with a branched molecular scaffold and one with a pericondensed, fused structure, \textbf{F1}.
These result in varied degrees of $\pi$-conjugation in the dyes, decreasing from \textbf{F1} to \textbf{B2}.
The electronic and spectroscopic characteristics of the dyes attached to a colloidal
TiO$_2$ nanocluster were determined by means of DFT and linear-response TDDFT calculations.
From the calculated absorption spectra, it is observed that the absorption maxima of the dyes 
in the visual region is red-shifted from \textbf{B1} to \textbf{F1}, which positively 
influences their sunlight harvesting properties.
The MO excitations from the TDDFT spectra unveil that transitions from the HOMO
are prevailing for the excited states in the visual region.
This confirms the usage of one-electron KS orbitals as the eigenstates of the system 
with one active electron in the HOMO as an adequate approximation.
In addition, the pDOS of the molecular orbitals for the dye, the substrate, 
and the entire device were computed to confirm their advantageous energetic alignment.
In essence, the results reveal that our model systems mimic the fundamental properties of realistic
DSSCs and are appropriate to simulate photoinduced charge migration processes.
To supplement the static considerations, the state injection times and currents are 
evaluated for each DSSC complex.
The parameter-free injection times are on the same time scale as the experimental 
value of $6\,{\rm fs}$ found for similar systems.
The state currents, which includes the state excitation probability, reveals that the 
\textbf{F1} dye yields the highest photocurrents followed by the \textbf{B1} chromophore.

Within a time-dependent single active electron approach (TDKS), the laser-induced 
charge migration process from the dye into the TiO$_2$-cluster in the model solar cell
is investigated in real time.
A broad band excitation of the visual absorption band initiates the electron dynamics
from the ground state.
The excited electron is injected into the substrate and absorbed by the oxygen atoms 
at the bottom of the cluster.
The population evolution exhibits a variability of the overall injection times, slightly 
increasing from \textbf{F1} to \textbf{B2}.
This is due to fluctuations between the ground state and a certain subset of excited states.
The mechanistic pathway of the electron injection is unveiled with electronic yields 
which are partitioned for specific molecular fragments.
This reveals an oscillation of electron density between the electron donor group 
to the acceptor group of the dye, with simultaneous injection into the titania cluster.
A detailed look at the evolution of the electronic yields shows increased 
fluctuation of electron density between the electron releasing and withdrawing moieties 
in \textbf{B1} and \textbf{B2}.
This observation is supported by time-dependent electron density plots at representative 
time steps.
The comparison between one of the branched dyes \textbf{B1} and the fused chromophore 
\textbf{F1} confirms the crucial influence of the different structural features on 
the charge migration process.
In particular, it seems that a more weakly electron donor group far away from the interface can play
a role in the injection dynamics.
Also, the degree of $\pi$-conjugation can be seen to improve the light harvesting properties
and the hybridization of the chromophore at the interface, thereby increasing the
photovoltaic efficiency of the device.

\section{Acknowledgment}

The authors thank funding from the Deutsche Forschungsgemeinschaft through 
project TR1109/2-1 and Hans-Christian Hege for providing the ZIBAmira visualization
program.
Generous computing resources from the Scientific Computing Services Unit of the Zentraleinrichtung
f{\"u}r Datenverarbeitung (ZEDAT) at Freie Universit{\"a}t Berlin are gratefully acknowledged.

\end{document}